\documentclass[twocolumn,prb,showpacs,floatfix]{revtex4}

\usepackage{graphicx}

\begin{document}

\title{Two Superconducting Phases in CeRh$_{1-x}$Ir$_x$In$_5$}

\author{M.~Nicklas}
 \altaffiliation{Present address: Max Planck Institute for Chemical Physics of
 Solids, N{\"o}thnitzer Str. 40, 01187 Dresden, Germany.}
\author{V.~A.~Sidorov}
 \altaffiliation{Permanent address: Institute for High Pressure
 Physics, Russian Academy of Sciences, Troitsk, Russia.}
\author{H.~A.~Borges}
 \altaffiliation{Permanent address: Pontif{\'i}cia Universidade Cat{\'o}lica
 Rio de Janeiro, Dept. F{\'i}sica, BR-22452-970 Rio De Janeiro, Brazil.}

\author{P.~G.~Pagliuso}
 \altaffiliation{Permanent address: Cidade Universitaria, BR-13083-970
 Campinas-SP, Brazil.}
\author{J.~L.~Sarrao}
\author{J.~D.~Thompson}

\affiliation {Los Alamos National Laboratory, Los Alamos, NM 87545.}

\date{\today}

\begin{abstract}
Pressure studies of CeRh$_{1-x}$Ir$_x$In$_5$ indicate two superconducting
phases as a function of $x$, one with $T_c \geq 2$ K for $x<0.9$ and the other
with $T_c<1.2$ K for $x>0.9$. The higher $T_c$ phase, phase-1, emerges in
proximity to an antiferromagnetic quantum-critical point; whereas, Cooper
pairing in the lower $T_c$ phase-2 is inferred to arise from fluctuations of a
yet to be found magnetic state. The $T$-$x$-$P$ phase diagram of
CeRh$_{1-x}$Ir$_x$In$_5$, though qualitatively similar, is distinctly
different from that of CeCu$_2$(Si$_{1-x}$Ge$_x$)$_2$.

\end{abstract}

\pacs{71.27.+a, 74.70.Tx, 74.62.Dh, 74.62.Fj}

\maketitle

As a conventional superconductor is cooled below $T_c$, a finite energy gap in
the electronic density-of-states $N(E_F)$ opens over the entire Fermi surface.
This gap to quasiparticle excitations produces an exponential temperature
dependence of physical properties that depend on $N(E_F)$, eg. specific heat,
thermal conductivity, and spin-lattice relaxation rate. In contrast, power-law
dependences of these properties are found in superconducting heavy-fermion
systems \cite{Grewe91} as well as in cuprates,\cite{Orenstein00} ruthenates
\cite{Ishida00} and low-dimensional organics.\cite{McKenzie97} The existence
of these power laws can be understood if the superconducting energy gap,
instead of being everywhere finite, is zero on parts of the Fermi surface so
that the excitation spectrum starts from zero energy. These qualitative
departures from conventional behavior suggest that Cooper pairing may be
mediated by excitations other than phonons. In each class of materials
mentioned above, a 'dome' of superconductivity emerges in proximity to a
magnetic transition that is tuned toward zero temperature by applied pressure
or changes in chemical composition. The close proximity to magnetism and
evidence for power-law behaviors below $T_c$ argue for magnetically mediated
superconductivity in which the orbital component of the superconducting order
parameter is greater than zero and power laws reflect the nodal structure of
the order parameter.\cite{Sigrist91}

With two notable counter examples, a single dome of superconductivity tends to
appear only in a relatively narrow range of tuning parameter values. One of
these counter examples is U$_{1-y}$Th$_y$Be$_{13}$. In this case, substitutions
of nonmagnetic Th for U cause a non-monotonic variation of $T_c(y)$ with a
sharp, non-zero minimum in $T_c$ near $y=0.019$ that is not due simply to
pair-breaking effects, since superconductivity persists to at least
$y=0.06$.\cite{Lambert86} Pressure studies\cite{Lambert86} of the $T_c(y)$
phase diagram reveal that the minimum in $T_c$ near $y=0.019$ evolves into a
range of $y$ where there is no superconductivity and provide convincing
evidence that the $T_c$ minimum at atmospheric pressure delineates two
distinct superconducting phases. Though weak magnetism coexists with
unconventional superconductivity for $0.019 <y < 0.042$ at atmospheric
pressure, the origin of two distinct transitions remains unclear.

The other counter example is the prototypical heavy-fermion compound
CeCu$_2$Si$_2$.\cite{Steglich79} Until recently, its inexplicably robust
superconductivity with respect to pressure and complex variation of $T_c(P)$
have appeared incompatible with magnetically mediated superconductivity.
Detailed pressure studies of CeCu$_2$Si$_2$ and its slightly larger volume
relatives CeCu$_2$(Si$_{1-x}$Ge$_x$)$_2$ reveal the existence of two distinct
domes of different superconducting phases, one at low pressures controlled by
proximity to an antiferromagnetic quantum-critical point and a second at
higher pressures that coincides with a weakly first-order phase boundary
delineating an isostructural volume collapse.\cite{Yuan2003} The former is
consistent with a magnetic pairing mechanism, whereas the latter suggests that
density fluctuations and associated Ce-valence fluctuations are involved in
Cooper pairing.

CeRh$_{1-x}$Ir$_x$In$_5$ is a candidate for demonstrating two superconducting
phases. CeRhIn$_5$ \cite{Hegger00} and CeIrIn$_5$ \cite{Petrovic01} are
isostructural, isovalent heavy-fermion compounds that form solid solutions in
which the ratio of tetragonal lattice parameters, $c/a$, varies linearly
across the series.\cite{Pagliuso01b} With progressive substitutions of Rh by
Ir in CeRh$_{1-x}$Ir$_x$In$_5$, the ground state at atmospheric pressure
evolves continuously, just as it does in CeRhIn$_5$ with applied pressure,
\cite{Mito03} from antiferromagnetic ($x < 0.3$) to antiferromagnetic with
coexisting superconductivity ($0.3 < x < 0.6$) and finally to superconducting
without apparent evidence for long-range magnetic order ($x >
0.6$).\cite{Pagliuso01b} As the end composition CeIrIn$_5$ is approached,
there is a cusp-like minimum in $T_c(x)$ near $x=0.9$ where bulk
superconductivity is suppressed. The specific heat anomaly at $T_c$ for this
composition is small, $\Delta C/\gamma T_c\approx 0.14$, which is only about
10\% of the weak-coupling BCS value,\cite{Bianchi01} and may be non-zero
because of slight variations in Rh/Ir concentrations throughout the sample.
Though suppression of bulk superconductivity with small additions of Rh in
CeIrIn$_5$ might arise from Cooper-pair breaking by non-magnetic Rh
'impurities', for $x
> 0.9$ or $x < 0.9$, the specific heat jump at $T_c$ is comparable to the BCS
value,\cite{Pagliuso01b,Bianchi01} and below $T_c$, the relaxation rate
$1/T_1\propto T^3$ and specific heat divided by temperature $C/T\propto T$,
indicative of unconventional superconductivity.\cite{Zheng01} As we will show,
the cusp in $T_c$ near $x=0.9$ in CeRh$_{1-x}$Ir$_x$In$_5$ evolves with applied
pressure to become a range of compositions that separates two superconducting
phases.

Simultaneous electrical resistivity and {\it ac} susceptibility measurements
were used to study the response to pressure of high quality single crystals of
CeRh$_{1-x}$Ir$_x$In$_5$ for $x=0$, 0.1, 0.25, 0.5, 0.75, 0.85 and 1.0. The
crystals, grown from excess In flux, were carefully screened at atmospheric
pressure by SQUID magnetometry to ensure the absence of free In. Pressures to
2.3 GPa were generated in a Be-Cu clamp-type cell with Flourinert as the
pressure-transmitting medium, and at least seven, approximately equally
spaced, pressure measurements were made on each composition. The inductively
measured shift in the superconducting transition of high purity Sn or Pb
determined the clamped pressure at low temperatures.

\begin{figure}[t]
\includegraphics[angle=0,width=65mm,clip]{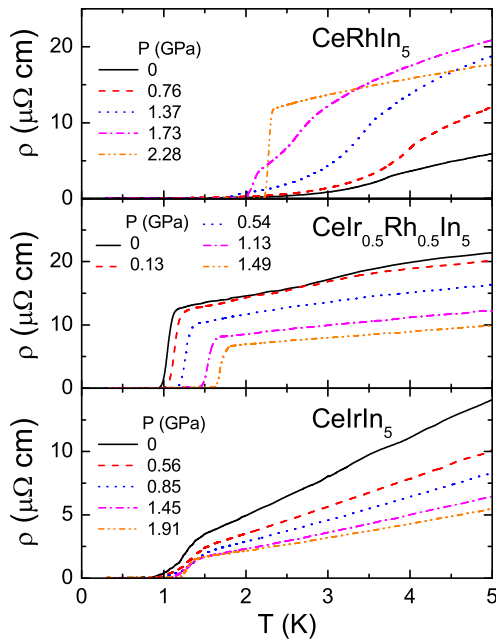}
\caption{\label{fig1} Resistivity versus temperature for three compositions of
CeRh$_{1-x}$Ir$_x$In$_5$ at representative pressures. Responses at other
values of $x$ are intermediate to those shown here. }
\end{figure}

Figure 1 shows the electrical resistivity at various pressures for $x=0$, 0.5
and 1.0 in CeRh$_{1-x}$Ir$_x$In$_5$. These responses are representative of the
series. For $x<0.5$, the low temperature resistivity increases initially with
applied pressure and the temperature $T_{max}$ at which the resistivity is a
maximum (not shown) decreases with $P$. Near and above $x=0.5$, opposite
trends appear --- the low temperature resistivity decreases and the
resistivity maximum moves to higher temperatures with applied pressure. As
seen in Fig.~1, Rh/Ir substitutions have a small effect on potential
scattering since the limiting resistivity just above either an
antiferromagnetic or superconducting phase transition at atmospheric or high
pressure varies from about ${\rm 2~\mu\Omega cm}$ for $x=0$ and 1.0 to about
${\rm 7~\mu\Omega cm}$ for $x=0.5$. Qualitatively, this reflects Nordheim's
rule for isovalent substitutions\cite{Olsen62} and is a further indication of
sample homogeneity. Superimposed on this frozen disorder scattering are
comparable or larger pressure-dependent changes in the inelastic scattering
rate. For $x<0.5$, pressure enhances the scattering rate as magnetic order is
replaced by superconductivity; whereas, for $x=0.5$, the scattering rate at
atmospheric pressure is already relatively large and decreases with applied
pressure and this trend continues with increasing $x$. The variation in the
low-temperature resistivity of this CeRh$_{1-x}$Ir$_x$In$_5$ series at
atmospheric pressure is analogous to responses found in several
antiferromagnets as they are tuned by applied pressure toward a quantum
critical point. \cite{Miyake02} This analogy argues that Ir substitution for Rh
acts principally as an effective applied pressure and that there is a
quantum-critical point at atmospheric pressure in the series near $x\geq 0.5$.
Indeed, the ambient-pressure N{\'e}el temperature drops to $T=0$ at
$x_c\approx0.65$ where the specific heat begins to diverge
logarithmically,\cite{Pagliuso01a} and, as shown in Fig.~2, Ir substitution
and applied pressure are demonstrably equivalent for $x\leq 0.25$. The rigid
shift by a constant pressure of the superconducting transition $T_c(P)$, the
N{\'e}el temperature $T_N(P)$ and the temperature $T_{max}(P)$, where the
resistivity is a maximum, scales each onto a common curve. For these three
compositions, the rigid pressure shift $P_r{\rm(GPa)}\approx\,10 x^2$, which,
extrapolating to $x=1$, implies that CeIrIn$_5$ is under an effective chemical
pressure of about 10 GPa relative to CeRhIn$_5$. This straightforward scaling
breaks down for $x>0.3$, indicating additional effects of Ir substitution.

\begin{figure}[t]
\includegraphics[angle=0,width=65mm,clip]{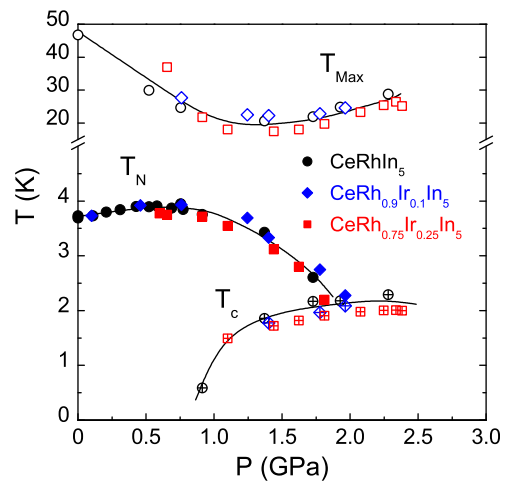}
\caption{\label{fig2} Pressure dependence of the temperature $T_{max}$ where
the resistivity is a maximum, the N{\'e}el temperature $T_N$ and
superconducting transition temperature $T_c$ for $x=0$ (circles), 0.1
(diamonds), and 0.25 (squares) in CeRh$_{1-x}$Ir$_x$In$_5$. Data, shifted by
constant pressures of 0, 0.1, and 0.6 GPa for $x=0$, 0.1 and 0.25,
respectively, scale onto common curves as shown.}
\end{figure}

\begin{figure}[t]
\includegraphics[angle=0,width=65mm,clip]{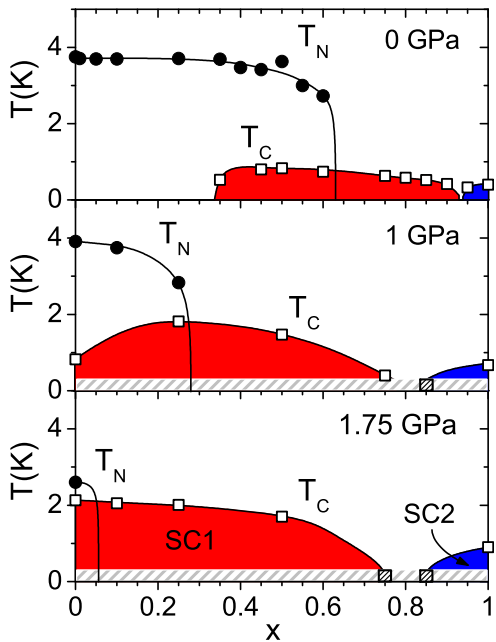}
\caption{\label{fig3} Representative $T-x$ phase diagrams for
CeRh$_{1-x}$Ir$_x$In$_5$ at $P=0$, 1.0 and 1.75 GPa. The cusp-like suppression
of $T_c$ near $x=0.9$ at $P=0$ evolves into a range of compositions where
$T_c<0.3$ K at higher pressures.  Plots of $T_c(P)$ for $x=0.75$ and 0.85 (not
shown explicitly) strongly suggest that $T_c=0$ for these compositions at 1.75
GPa and higher pressures. SC1: phase-1 superconductivity; SC2: phase-2
superconductivity.}
\end{figure}

Linear interpolations of $T_c(P)$, defined by the onset of a diamagnetic
response in {\it ac} susceptibility, and $T_N(P)$, determined from a change in
slope of $\rho(T)$, for each value of $x$ allow the construction of isobaric
$T-x$ phase diagrams. Data in the upper panel of Fig.~3 are results from
ambient-pressure measurements,\cite{Pagliuso01b} and those in the middle and
lower panels are representative $T-x$ diagrams at pressures of 1.0 and 1.75
GPa. Similar isobaric diagrams at intermediate pressures confirm the smooth
evolution seen in Fig.~3, and, in particular, the $T-x$ phase diagram at 2 GPa
shows no evidence for antiferromagnetism. As seen in Fig.~3, the cusp-like
suppression of $T_c$ near $x=0.9$ at $P=0$ evolves with applied pressure to
become a range of compositions $0.75\leq x\leq 0.85$ where no bulk
superconductivity is detected above 0.3 K by {\it ac} susceptibility.
Therefore, in CeRh$_{1-x}$Ir$_x$In$_5$ under pressure, there is a range of
compositions separating two superconducting phases, phase-1 with $T_c\geq  2$
K for $x<0.75$ and phase-2 with $T_c<1.2$ K for $x>0.85$.\cite{footnote}

The results of Fig.~3 appear analogous to the evolution of $T_c(y,P)$ in
U$_{1-y}$Th$_y$Be$_{13}$, particularly, if we consider CeRh$_{1-x}$Ir$_x$In$_5$
as Rh-doped CeIrIn$_5$,  as well as to the observation of two superconducting
phases in CeCu$_2$(Si,Ge)$_2$. In the latter, each dome of superconductivity
is controlled by proximity to a distinctly different transition that is tuned
to $T\rightarrow0$ by pressure. This conclusion was possible by realizing that
Ge substitution for Si expands the unit-cell volume and that this expansion
can be compensated by an externally applied pressure to produce nearly
identical superconducting phase diagrams as a function of cell volume for both
CeCu$_2$Si$_2$ and CeCu$_2$Ge$_2$.\cite{Yuan2003} A similar argument is
inferred from the pressure scaling shown in Fig.~2. If the primary role of Ir
substitutions for Rh is to decrease the cell volume, then the observation of
two superconducting phases in CeRh$_{1-x}$Ir$_x$In$_5$ suggests that a second
superconducting phase also might emerge in CeRhIn$_5$ at much higher pressures
than investigated here. Besides a dome of superconductivity centered near the
antiferromagnetic critical point at $P_c\approx 2.5$ GPa where $T_c$ exceeds 2
K, Muramatsu {\it et al.}\cite{Muramatsu01} have reported a second dome of
superconductivity in CeRhIn$_5$ with a maximum $T_c\approx 1$ K near 6.5 GPa.
Considering that details of Ir/Rh substitution were ignored in estimating the
effective pressure in CeIrIn$_5$, this estimate and the observed pressure of
6.5 GPa are in good agreement and further suggest that the second,
high-pressure dome of superconductivity in CeRhIn$_5$ is analogous to phase-2
superconductivity in CeRh$_{1-x}$Ir$_x$In$_5$.

This simple volume-based extrapolation was implied from the empirical
observation that $P_r \propto x^2$ for $x\leq 0.25$.  Studies at atmospheric
pressure show, however, that $T_c$'s of CeRh$_{1-x}$Ir$_x$In$_5$ are a linear
function of the ratio of tetragonal lattice parameters $c/a$ and not cell
volume ($a^2c$). \cite{Pagliuso02} This apparent dichotomy suggests that $c/a$
is not a monotonic function of pressure even though the cell volume is.
Pressure-dependent structural studies of CeRhIn$_5$ confirm this
suggestion:\cite{Kumar04} $c/a$ exhibits two maxima as a function of pressure,
one near 2.5 GPa and a second near 6 GPa. The correspondence between these
maxima and those in $T_c(P)$ reinforces the relationship between $T_c$ and
$c/a$ found in the Rh/Ir solid solutions at atmospheric pressure. The
pronounced non-monotonic variation of $c/a(P)$ in CeRhIn$_5$, though not
directly established in other members of CeRh$_{1-x}$Ir$_x$In$_5$ or in
isostructural CeCoIn$_5$, also may account for the different responses of $T_c$
to uniaxial pressure observed in CeIrIn$_5$ and CeCoIn$_5$.\cite{Oeschler03}

Though the emergence of two superconducting phases in CeRh$_{1-x}$Ir$_x$In$_5$
and CeRhIn$_5$ under pressure appears similar to the non-monotonic variation
of $T_c(P)$ in CeCu$_2$(Si,Ge)$_2$, there is an important distinction. In the
latter, there are well-defined regimes of pressure where $T_c<1$ K and $T_c>
2$ K, but, the high-pressure, high-$T_c$ regime is accompanied by topological
changes in the Fermi surface\cite{Thomas96} and/or an increase in ground state
degeneracy\cite{Jaccard99,Bellarbi84} so that superconductivity with different
$T_c$'s develops out of qualitatively different electronic states. This is not
true in CeRh$_{1-x}$Ir$_x$In$_5$ and CeRhIn$_5$ under pressure. deHaas-van
Alphen studies find that, except for expected quantitative changes due to
their slightly different ratio of tetragonal lattice parameters, CeIrIn$_5$ at
$P=0$ and superconducting CeRhIn$_5$ $(P >0)$ have the same Fermi-surface
topology and comparably large quasiparticle masses.\cite{Haga01,Shishido02}
Further, at atmospheric pressure, the electronic entropy to 5 K is
$(0.5\pm0.05)R\ln 2$ for all $x$,\cite{Pagliuso01b} indicating the same ground
state degeneracy.

On the basis of scaling shown in Fig. 2, we assume reasonably that
superconductivity in phase-1 has the same origin as in CeRhIn$_5$ near and
below 2.5 GPa, namely that superconductivity is mediated by excitations
associated with proximity to an antiferromagnetic quantum-critical point. The
pairing mechanism for phase-2 superconductivity is not so obvious since
antiferromagnetic order appears to be well removed from this part of the phase
diagram and there is no evidence for a line of valence transitions as a
function of $x$ or $P$.  Like superconductivity in phase-1 where $C/T\propto
T$ and $1/T_1\propto T^3$ below $T_c$, the same power laws are
found\cite{Thompson03} in CeIrIn$_5$, which is representative of phase-2
superconductivity, and indicate an unconventional mechanism for
superconductivity in phase-2. The pairing mechanism for phase-2 is suggested
from thermal expansion measurements on CeIrIn$_5$ in a field sufficient to
destroy bulk superconductivity. In these experiments, the coefficient of
$c$-axis thermal expansion $\alpha_c= a\,T^{0.5} +b\,T$, a temperature
dependence expected for thermal expansion dominated by three-dimensional
critical fluctuations at an antiferromagnetic quantum-critical
point.\cite{Oeschler03} These observations, together with a non-Fermi-liquid
like $1/T_1$ above $T_c$ in CeIrIn$_5$,\cite{Zheng01,Kohori01} imply that
phase-2 superconductivity in CeRh$_{1-x}$Ir$_x$In$_5$ for $x > 0.85$ and, by
inference, in CeRhIn$_5$ at $P> 5$ GPa is mediated by fluctuations arising
from some form of hidden magnetic order. One possibility is that this hidden
order manifests itself in field-induced magnetic transitions observed in
CeIrIn$_5$ near 40~T\,\cite{Stewart02} and in CeRhIn$_5$ near
50~T.\cite{takeuchi01} Whatever the precise nature of this hidden magnetic
order, the lower $T_c$ of phase-2 superconductivity suggests that
pair-mediating fluctuation spectrum is more nearly 3-dimensional, coupling
electronic states less efficiently than magnetic excitations associated with
$T_N$ and phase-1 superconductivity.

We thank A. V. Balatsky, Y. Bang, C. D. Batista, Z. Fisk, and M. J. Graf for
useful discussions. Work at Los Alamos was performed under the auspices of the
U.S. Department of Energy. V. A. S. acknowledges support of the Russian
Foundation for Basic Research, Grant No. 03-02-17119.


\begin{thebibliography}{100}
\bibitem{footnote} Additional measurements find that $T_c$ is less than 1 K for CeRh$_{0.5}$Ir$_{0.5}$In$_5$
at $P=2.1$ GPa and that the bulk superconducting transition of CeIrIn$_5$ does
not exceed 1.2 K at pressures to $P=4$ GPa.
\end{thebibliography}
\end{document}